# Multi-Angle Beam Scanning Transmission Electron Microscopy for Diffraction-Contrast-Free Magnetic Imaging

*Akira Yasuhara*[1], Zentaro Akase[2] and Takumi Sannomiya*[3]*

[1] JEOL Ltd. 3-1-2 Musashino, Akishima, Tokyo, 196-8558, Japan.

[2] Division of Materials Science, Nara Institute of Science and Technology, 8916-5 Takayama, Ikoma, Nara 630-0192, Japan

[3] Department of Materials Science and Engineering, School of Materials and Chemical Technologies, Institute of Science Tokyo, 4259 Nagatsuta, Midoriku, Yokohama, 226-8501, Japan.

**Corresponding Authors**

* Akira Yasuhara (Email: ayasuhar@jeol.co.jp)

* Takumi Sannomiya (Email: sannomiya@mct.isct.ac.jp)

ABSTRACT

Recent development of the differential phase contrast (DPC) scanning transmission electron microscopy (STEM) technique has significantly advanced materials science supported by high-sensitivity detectors and powerful data processing. Especially the 4D-STEM technique, employing a pixelated camera as a STEM detector, enables extensive information acquisition through post-processing of large datasets. However, magnetic or electric field imaging using DPC is often disturbed by diffraction contrasts intrinsic to the crystalline material. In this paper, we demonstrate elimination of the diffraction contrasts for magnetic field imaging by using angularly distributed multiple electron beams generated by a multi-hole aperture within the 4D-STEM framework. This multi-angle beam configuration suppresses irrelevant crystalline contrast and yields direct magnetic information with high sensitivity from the specimen. The resulting 4D-STEM datasets also enable additional analysis of sub-beam size magnetic domain structures, domain walls, and diffraction-related information.

## I. INTRODUCTION

Direct observation of magnetic domains using (scanning) transmission electron microscopy ((S)TEM) has been widely applied to magnetic materials investigation, which offers high spatial resolution based on the electron probe as well as simultaneous acquisition of the structural and elemental information. Especially, differential phase contrast (DPC) imaging, which detects beam deflection arising from Lorentz forces and Coulomb potentials within the specimen,[1,2] has recently attracted considerable attention owing to the development of spherical-aberration correctors that enable sub-angstrom, high-brightness probes, together with high-sensitivity segmented detectors comprising multiple detection elements.[3] While DPC contrast is highly sensitive to the electromagnetic fields of specimens, the signal is strongly influenced by diffraction effects from the crystalline structure. This diffraction contrast in the DPC image can often complicate the interpretation of the specimen's electromagnetic fields.[4,5] To mitigate the diffraction-induced contrast, a tilt-averaged STEM technique has been developed for magnetic-field-free microscopes[6,7] This approach reduces diffraction contributions, and several successful applications have been reported.[8,9] Nevertheless, its implementation requires dedicated pre- and post-specimen tilt/de-tilt coils, as well as high speed electrical circuitry and control systems for beam manipulation, so that the incident beam is tilted and subsequently re-aligned to the optical axis after transmitting the sample while spatially scanning the electron beam. And even with tilt-averaging, diffraction effects can remain, motivating different strategies.

Possible alternatives include the four-dimensional (4D) STEM framework, where angular pattern data (beam tilt information) are acquired using a pixelated detector while the electron probe is scanned across each position of the specimen forming a four-dimensional data cube. Not only simple DPCs[10-12] but also ptychographic approaches can indeed nicely retrieve the phase information.[13-15] Complementary to such large-data processing approaches, beam-shaping techniques in TEM, employing specialized apertures fabricated via MEMS or micro-fabrication in place of the conventional circular aperture on the beam axis, have attracted considerable interest especially for the electron phase manipulation or phase imaging. These techniques enable e.g. modulating the electron beams from planar to vortex[16,17]or visualizing phase contrasts for biological and soft materials.[18] Building on this capability, tailoring the probe by beam shaping can further enhance 4D-STEM-based analyses.

In this paper, we propose a multi-angle beam (MaB) STEM technique, where structured illumination and 4D detection are combined, for DPC-based magnetic field observation with highly suppressed diffraction contrast. We employ a multi-hole aperture (Fig. 1) allowing several angularly tilted electron beams simultaneously illuminated and scanned over the specimen, and the corresponding angular patterns, deflected by the Lorentz force within the sample, are recorded on a pixelated detector. Such a multi-hole aperture was previously utilized for compressive multi-beam STEM where a spatially multiplexed electron beam enabled super resolution imaging from a significantly down-sampled data even with a bucket detection.[19] Here, instead of multiplexing in the real space, we take advantage of multiplexing in the angular space (without "tilt-scanning") to realize diffraction-contrast-free DPC imaging for magnetic imaging.

While previously proposed tilt-scanning approaches require dedicated hardware and modification to the electron microscope column,[6-9] the proposed MaB-STEM method can be implemented on a conventional 4D-STEM system simply by replacing the condenser aperture with a multi-hole aperture. This allows angular averaging of diffraction-related contrast while preserving magnetic information, providing a practical and readily implementable solution for diffraction-contrast suppression in DPC imaging. Furthermore, the multi-hole aperture can increase the DPC contrast by providing a larger total aperture-edge length, thereby enhancing sensitivity to beam deflections induced by magnetic fields.

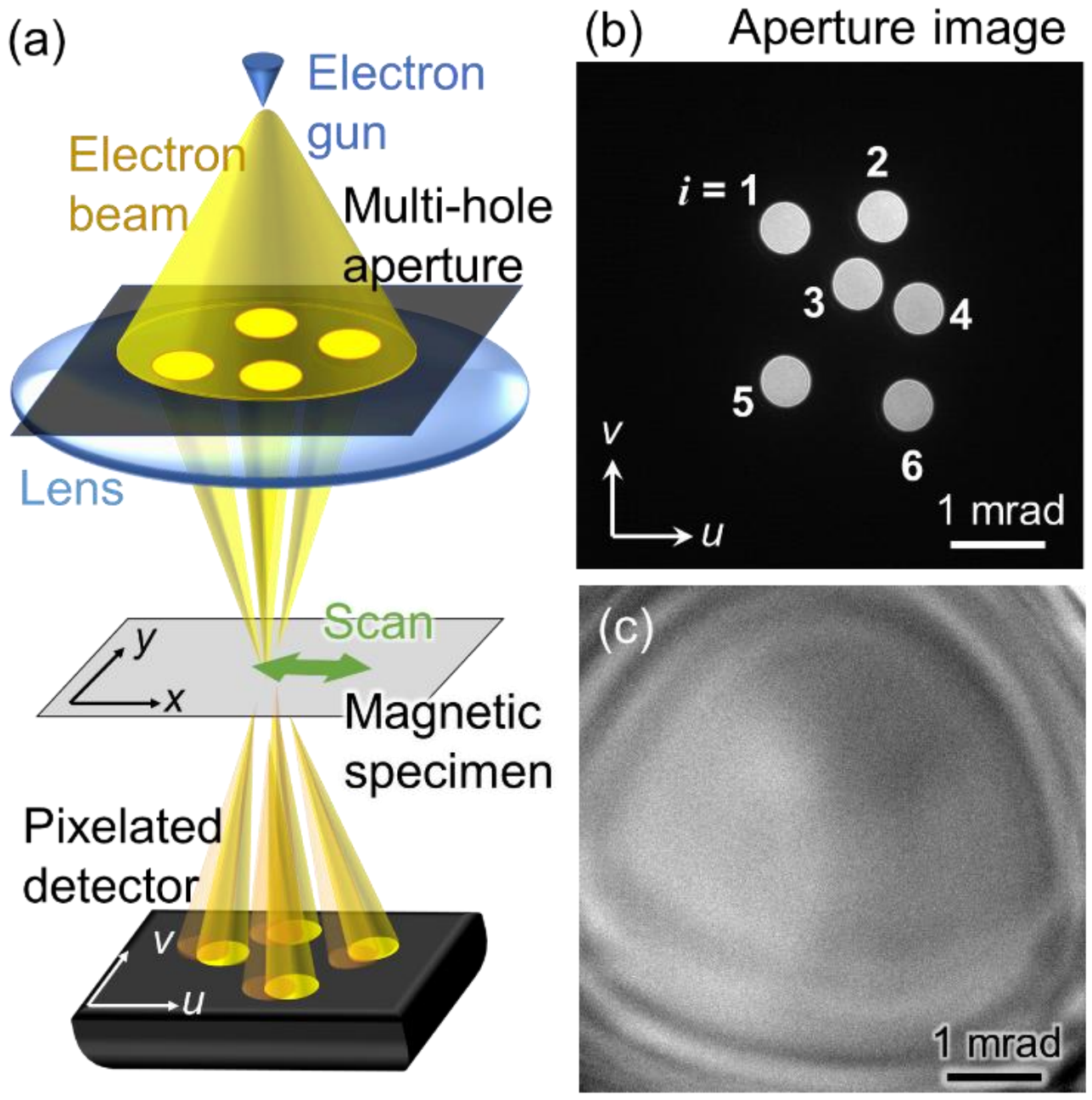

**Fig. 1.** (a) Schematic diagram of multi-angle beam (MaB) STEM. The arrows $\boldsymbol{x}$ and $\boldsymbol{y}$ denote the probe-scan coordinates on the specimen plane, whereas $\boldsymbol{u}$ and $\boldsymbol{v}$ denote the reciprocal-space coordinates on the detector plane. (b) Angular image of the multi-hole aperture installed on the condenser lens (CL) aperture position. The structure is fabricated by focused ion beam (FIB) method. The multi-hole aperture consists of six holes, numbered from $i = 1$ to 6. (c) Corresponding Ronchigram recorded under the same optical condition as panel b but without an aperture on an amorphous sample.

## II. EXPERIMENTAL METHODS

### A. Sample fabrication

As a test specimen, a soft-magnetic Mn–Zn ferrite material, which has an initial permeability of ~10,000 and has a cubic spinel structure composed of an iron-based oxide, was employed. The specimen consists of polycrystalline grains with random crystallographic orientations. To prepare a TEM sample, a thin lamella was prepared by a focused ion beam (FIB) method using an Helios NanoLab 600i (FEI, Thermo Fisher Scientific Inc., USA). First, a Mn–Zn ferrite block was extracted from the bulk materials using an in-situ micromanipulator in the FIB instrument and attached to a thin-bar TEM grid by Pt deposition. Subsequently, a thin lamella less than 100 nm thick was prepared using a Ga-ion beam.

### B. Instrumentation

A multi-hole aperture, designed to produce angularly distributed electron beams on the specimen plane, is set at the condenser aperture position in the TEM column.(Fig. 1a) Six randomly positioned holes, each with a diameter of 7 μm, as shown in Fig. 1b, were fabricated by FIB milling on a free-standing 10 μm-thick gold film. The random distribution of the holes is essential to avoid capturing systematic diffraction information with a certain symmetry. The six holes are numbered as #1-6, as shown in Fig. 1b, which we use in the analysis later. This aperture was installed in a JEM-ARM200F TEM/STEM (JEOL Ltd., Japan) equipped with a OneView CMOS camera (Gatan Inc., USA) as a pixelated detector. To analyze the intrinsic magnetic domains of the sample, the objective lens of the TEM was turned off to eliminate the strong magnetic field, which would otherwise saturate the specimen's magnetization along the electron beam direction. In the STEM imaging condition at an acceleration voltage 200 kV, the electron beam phase over the six openings of the aperture are approximately flat (least effect from geometrical aberrations), as shown in the Ronchigram of Fig. 1c, meaning that the six electron beams with different illumination angles are focused on the same position on the specimen. The flat region of the Ronchigram extends to $\pm 3$ mrad in radius around the optical axis. The maximum illumination tilt angle is approximately 2.5 mrad, well within the flat region of the Ronchigram (Fig. 1). The electron probe current and the convergence semi-angle of each circular aperture were 15 pA and 250 μrad, respectively. The diameter of each probe formed by 250 μrad semi-angle illumination is ~5 nm. Additional experimental parameters are summarized in Table S1 in the Supplementary Material. 4D-STEM data were acquired using STEM-X modules (Gatan Inc., USA), and post-processing analysis was performed using custom-developed Matlab programs.

## III. RESULTS AND DISCUSSION

### A. Appearance of diffraction contrast

Figure 2a shows a bright field (BF) STEM image constructed by integrating the entire transmission intensity. From the 4D-STEM dataset, two-dimensional angular pattern in the reciprocal space (***u***, ***v***) can be extracted at a given probe position on the specimen (***x***, ***y***). Representative angular patterns are shown in Fig. 2b, which are extracted from positions A–D indicated by colored circles in the BF-STEM image of Fig. 2a. The ***u*** and ***v*** arrows in Fig. 2a are shown to correspond them to the deflection direction in the angular (aperture) patterns in Fig. 2b. The intensity distribution of the aperture disk pattern varies significantly depending on the beam position on the specimen: At the contour position in the BF image, i.e. position A and D, the corresponding angular patterns in Fig. 2b shows severe intensity inhomogeneity among the disks. However, in the off-contour positions (B, C), the intensities of the six disks are homogeneously distributed, showing only the lateral shift of the entire pattern due to the magnetic field (flux) in the specimen.

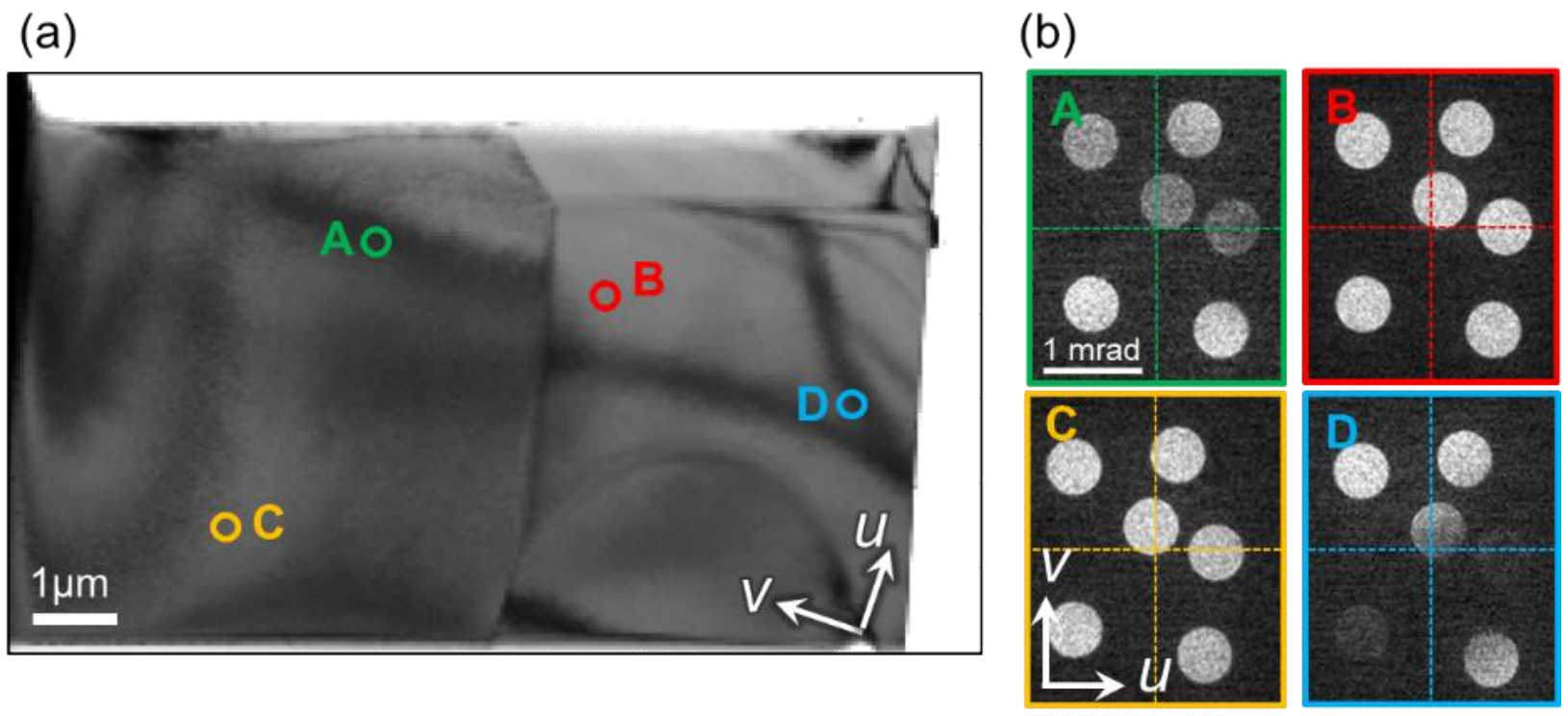

**Fig. 2** (a) STEM bright-field (BF) image of a Mn–Zn ferrite by integrating the entire angular pattern intensity. (b) Multi-hole aperture images extracted from the A–D positions marked by colored circles in panel a.

## B. DPC imaging by center-of-mass (CoM)

To construct the magnetization mapping from the 4D dataset, we firstly test the center of mass (CoM) analysis (see also Fig. S1 in Supplementary Material), which roughly corresponds to the differential signal of segmented detectors and is commonly used in STEM-DPC studies.[3,20,] Since each angular pattern in the 4D dataset consists of six aperture disks, as numbered in Fig. 1a, we performed CoM mapping of each disk, as shown in Fig. 3. The left column (Fig. 3a) shows the CoM shift maps of the ***u*** direction, and the right those of the ***v*** direction. Although magnetic contrasts with magnetic domains of a few micrometers (e.g. blue area at the bottom of Fig.3a panels), are visible, which is characteristic for soft magnetic materials like Mn-Zn ferrite, diffraction contrasts corresponding to the contours in the BF images in Fig. 2a are also very clearly overlaid, as indicated by white arrows. The influence of diffraction contrast on the CoM signal is schematically illustrated in Fig. S1 in Supplementary Material. Comparing the six disk mappings, the diffraction contrast varies depending on the disk, i.e. the tilt condition. These disk-dependent diffraction contrasts are also recognized in the corresponding STEM-BF images of individual disks, which are shown in Fig. S2 in the supplementary material.

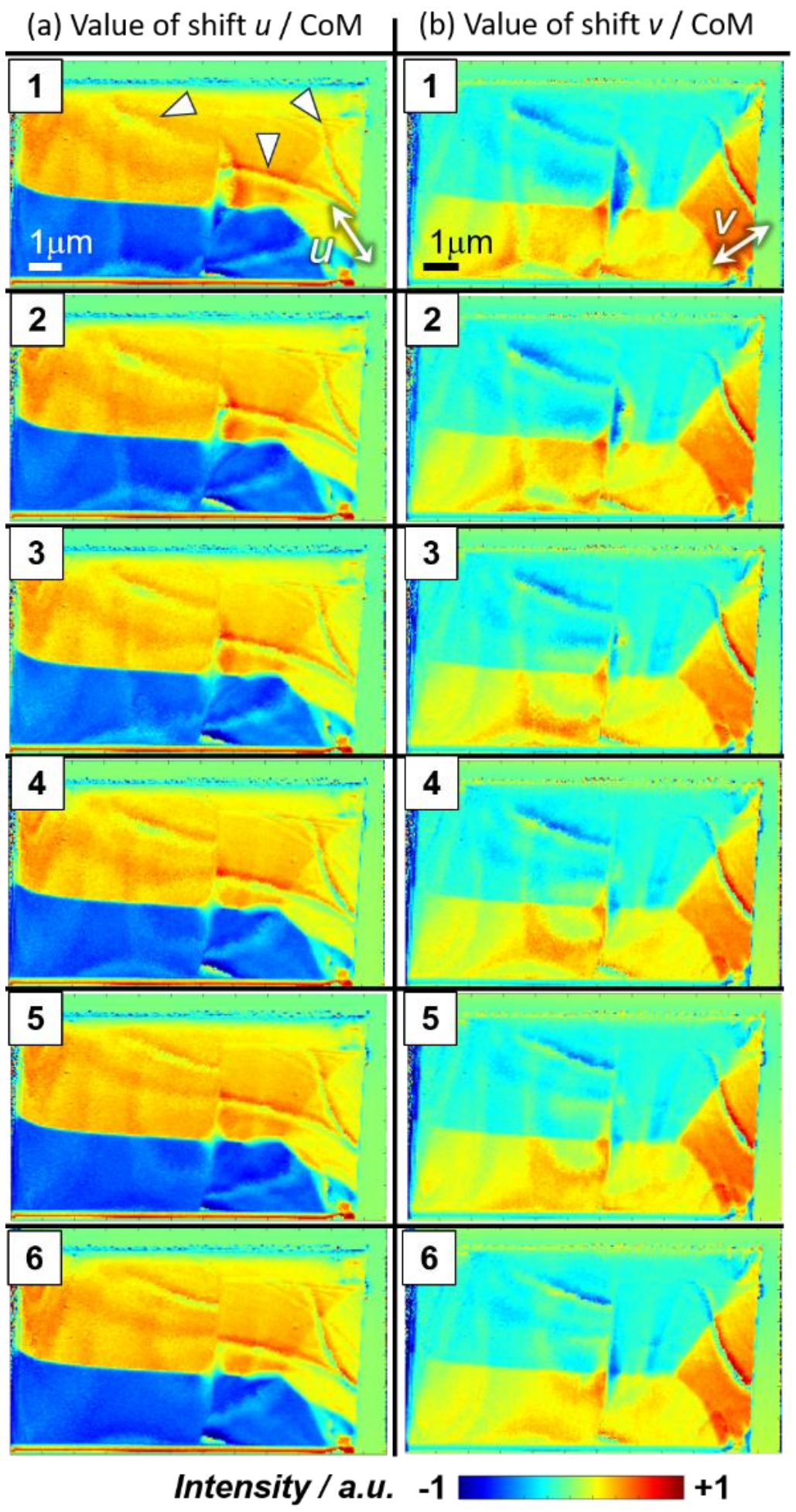

**Fig. 3.** DPC images obtained as center-of-mass (CoM) shifts for the six individual aperture disks. The numbers shown in the figure correspond to the aperture disk (hole) numbers in the angular pattern of Fig. 1b. The CoM shift maps along the ***u*** and ***v*** directions in the angular pattern are presented in the left (a) and right columns (b), respectively. Diffraction-related contour contrasts are marked by white arrows.

It is possible to reduce the diffraction contrast by averaging all the six magnetization maps, based on the fact that the diffraction contrast position is slightly different in each disk due to the different illumination angle. To suppress the diffraction contrast, we employed two different averaging approaches, as illustrated in Fig. S3(see Supplementary Material). In the first approach (Method 1 in Fig. S3), the average of the six maps of Fig. 3 is shown in Fig. 4a and b. Apparently the diffraction contrast contours are smoothed out and less visible and the magnetic contrasts including domains are more distinguishable. In the second approach (Method 2 in Fig. S3), the extracted six aperture disks can be superposed and averaged first, and the CoM shift can be calculated from the averaged single disk. The results of this “disk-average-first” approach are shown in Fig. 4c and d. Compared with the average of the six CoM shift maps (Fig. 4a, b), the diffraction contrasts indicated by white arrows are significantly reduced in the disk-average results (Fig 4c, d). This reduction of the diffraction contrast by disk-averaging (Fig. 4c, d) is because the signal intensity of the disks with large diffraction contrasts are weaker, and thus their contributions (or weights) become smaller by averaging the six disk intensities. However, the simple average of the maps (Fig. 4a, b) have equal contributions (weights) of all the disks regardless of the intensity in the disk, relatively “enhancing” the diffraction contrasts. The signal comparison of the single disk and the averaging is discussed later.

For another comparison, we also tested constructing CoM shift maps using the entire aperture patterns including all six disks, which are shown in Fig. 4e, f. In this case, magnetization information is hardly visible because of extremely enhanced diffraction contrasts. This strongly enhanced diffraction contrast is due to the inhomogeneous intensity distribution within the entire six disk patterns, as shown in Fig. 2b-A and D, which significantly shift the CoM even when the disks do not shift. These results shown here highlight the importance of utilizing appropriate angular signals to appropriately retrieve magnetization information from the specimen. We note that the "electrical tilt-scan" DPC method[6] corresponds to the latter "disk-average-first" scheme (Fig. 4c, d).

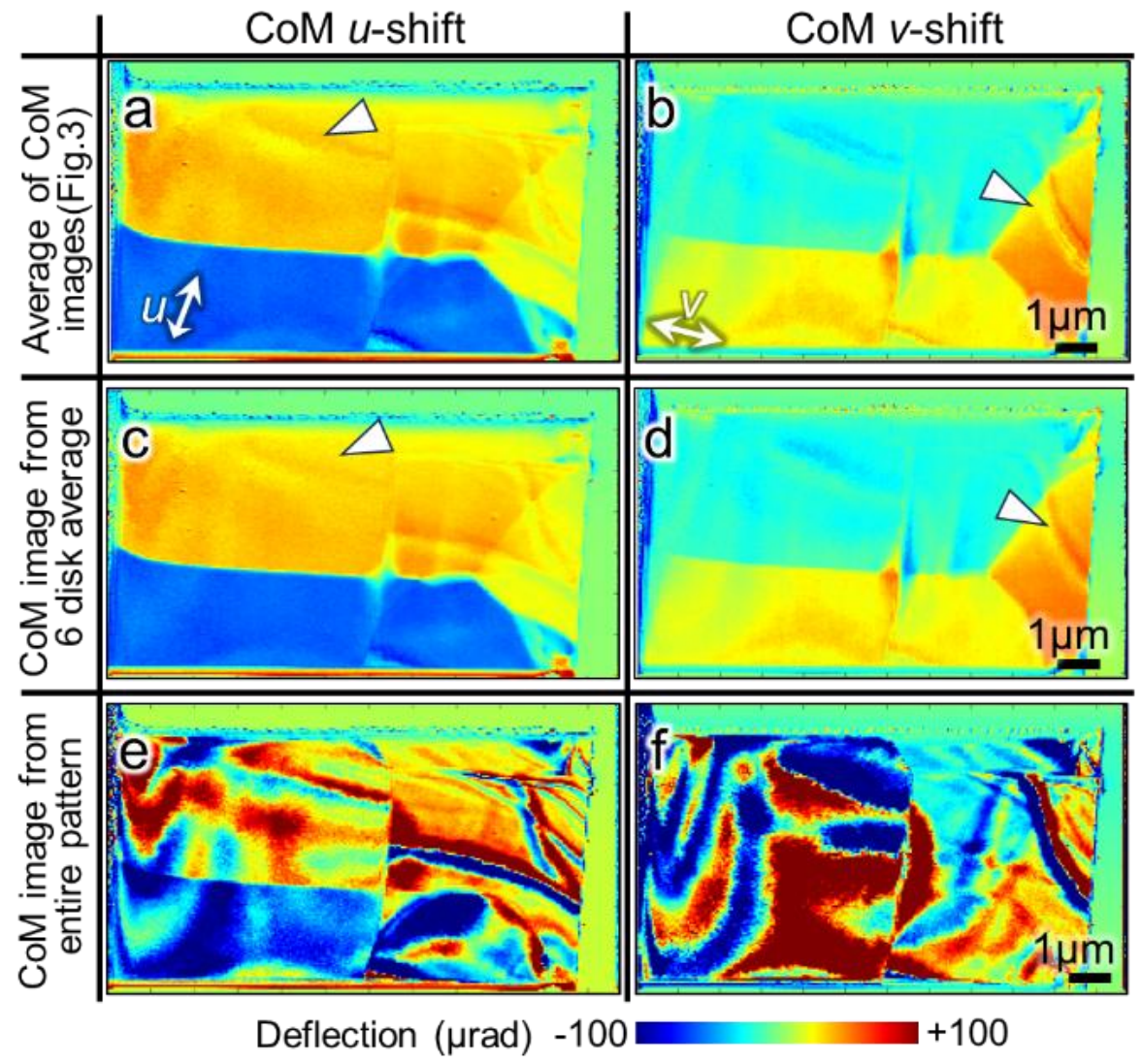

**Fig. 4.** (a, b) Average of six CoM shift maps shown in Fig. 3. (c, d) CoM shift maps computed after averaging the intensities by superposing the six extracted aperture disks at each electron beam position. (e, f) CoM shift maps calculated from the entire aperture pattern. The panels a, c, and e correspond to the CoM shift in ***u*** direction and b, d, and f to the CoM shift in ***v*** direction. White arrows indicate representative diffraction-related contour contrasts.

## C. DPC imaging by cross-correlation (CC)

As the second approach to evaluate the shift of electron beam in the angular space, we reconstructed the DPC images by cross-correlation (CC), which has been reported being more effective than the CoM approach.[12] For this CC method, a reference angular pattern is obtained from a vacuum region, where no beam deflection by the magnetization of the specimen is expected. The angular pattern of each electron beam position is then compared with this reference by CC and the shift of the CC peak is extracted by fitting it with a paraboloid.(see more details Fig. S4 in the Supplementary Material) The reproduced DPC images by CC of individual disks are shown in Fig. 5. Compared to the previous CoM results in Fig. 4, the CC shift maps exhibit better magnetization contrasts with less influences of the diffraction contrast. This signal improvement originates primarily from the sensitivity to the shift of the edge of disks in CC rather than the entire disk intensity, which is more influential in CoM. Since the diffraction causes the intensity distribution change but does not shift the disk, the CC method is less sensitive to the diffraction contrast than the CoM method, resulting in a clearer representation of magnetization with reduced crystalline diffraction effects.

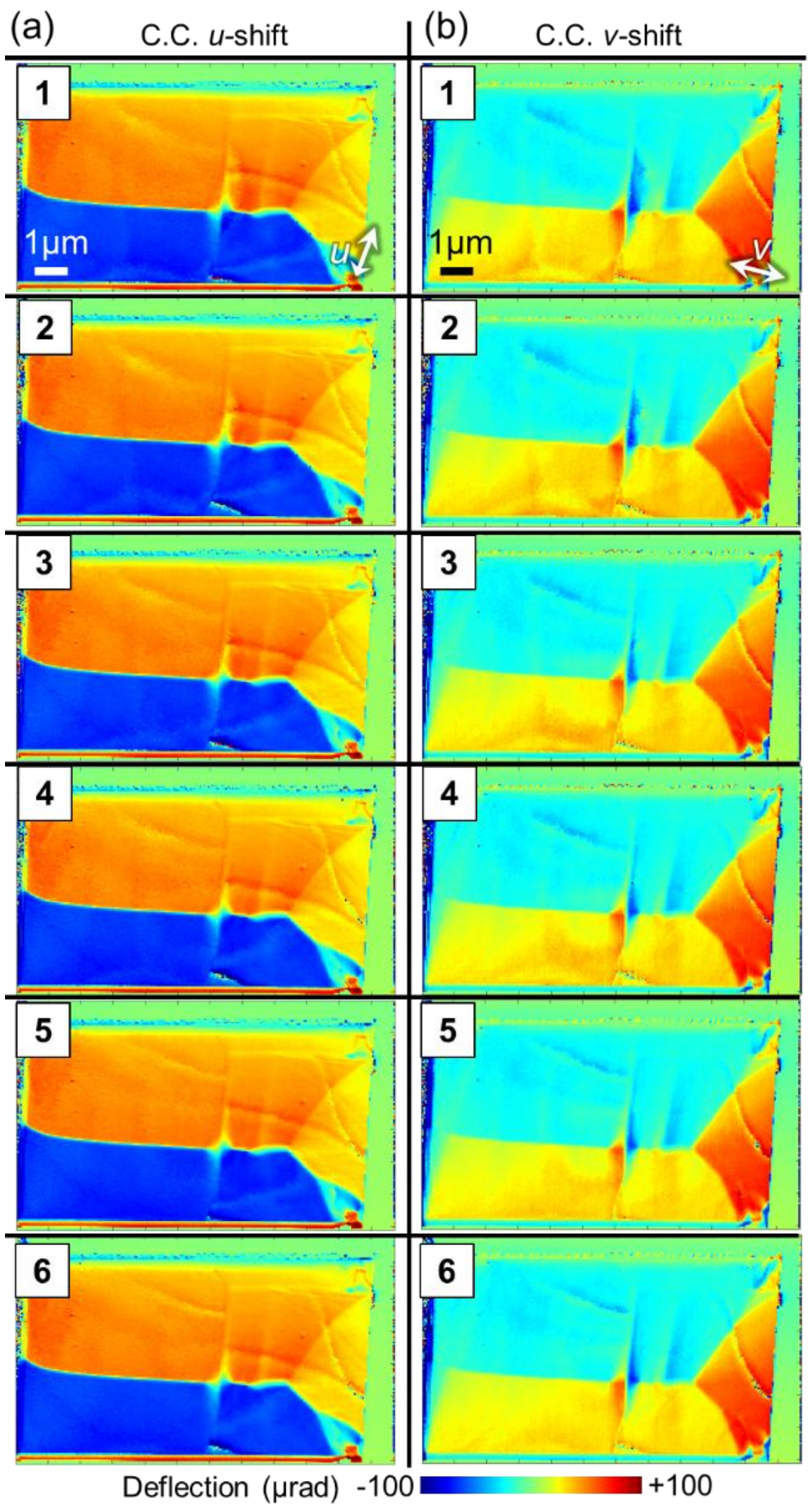

**Fig. 5.** DPC images obtained as CC peak shifts for the six individual aperture disks. The numbers correspond to the aperture disk (hole) numbers in the angular pattern of Fig. 1b. The CC peak shifts are along (a) ***u*** and (b) ***v*** directions in the angular patterns, which are indicated by arrows in the top most image (#1).

Figures 6a and b show averages of six DPC images from six disks of Fig. 5. The diffraction contrasts are further reduced by averaging because the diffraction contours appear at different position with different illumination angles. For the CC approach, superposing the disk patterns (similar to Fig. 4c, d) is equivalent to acquiring CC of the entire pattern. The DPC images by CC of the entire angular pattern are shown in Fig. 6c, d. This “entire-pattern” CC approach even more effectively suppresses the diffraction contrast compared with the average of six images of Fig. 6a, b, which are especially clear at the positions indicated by white triangular arrows. This improvement is essentially the same as the effect of disk averaging for the CoM approach in Fig. 4. In the entire-pattern CC analysis, aperture disks with lower intensity contribute less to the cross-correlation signal. Since the diffraction disks with strong diffraction contrasts generally exhibit reduced intensity, their contribution to the reconstructed image is weighted less, resulting in reduced diffraction-related contrast. Since the CC approach is sensitive to the “edges” of the aperture, these results here suggest that increasing the number of illumination holes (or just structuring the aperture more in a complex manner) should improve the magnetic signal.

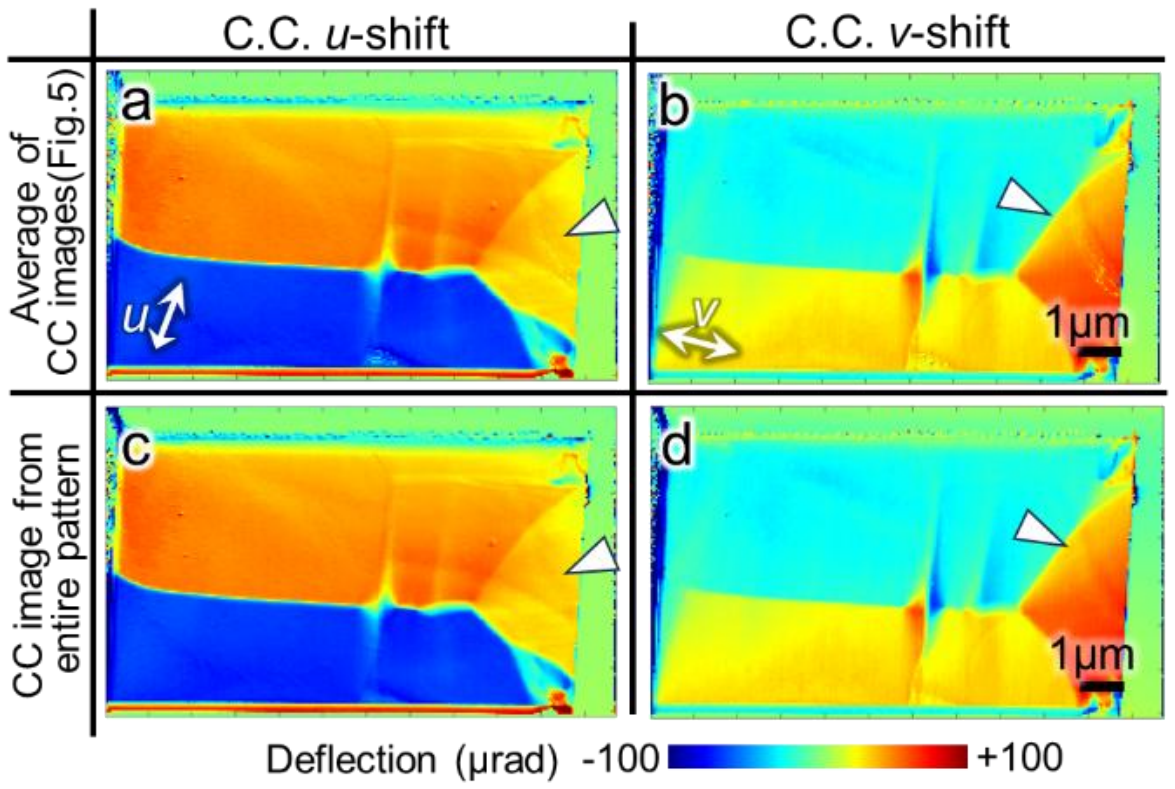

**Fig. 6** (a, b) Average of six CC peak shift maps shown in Fig. 5. (c, d) CC peak shift maps of the entire angular pattern including six disks. The panels a and c correspond to the shift in $\boldsymbol{u}$ direction, and panels b and d, to the shift in $\boldsymbol{v}$ direction. White triangular arrows indicate representative diffraction-related contour contrasts. Compared with the averaged CC maps (a, b), these contrasts are significantly suppressed in the entire-pattern CC maps (c, d).

### D. Toward Fully Diffraction-Contrast-Free Imaging: Edge Detection

As discussed above, the magnetic (Lorentz) deflection rigidly shifts the whole transmitted aperture pattern, and this shift is encoded mainly at the disk edges, which are defined by the aperture holes.[12] In contrast, diffraction contrast redistributes the intensity within the disks without moving their edges. Extracting only the edge positions should therefore further suppress the diffraction contrast. To this end, the edge position of each disk was detected at 96 positions along each disk rim, and normalized to its local intensity inside and outside the disk to retain the sub-pixel information. The shift of the disk was obtained by a least-squares fit. Unlike the CC analysis of the raw patterns (Fig. 6), this approach is largely insensitive to intensity variations caused by crystalline diffraction. The resulting shift maps are shown in Fig. 7. The diffraction contrast, such as bend contours, is further suppressed compared with the CC images in Fig. 6. However, the edge extraction uses only the pixels near the disk edges and amplifies high-frequency noise, so the maps show slightly more speckle-like noise. Further advanced image processing could improve the signal-to-noise ratio while keeping the diffraction contrast suppressed, however at the cost of increased computational loads.

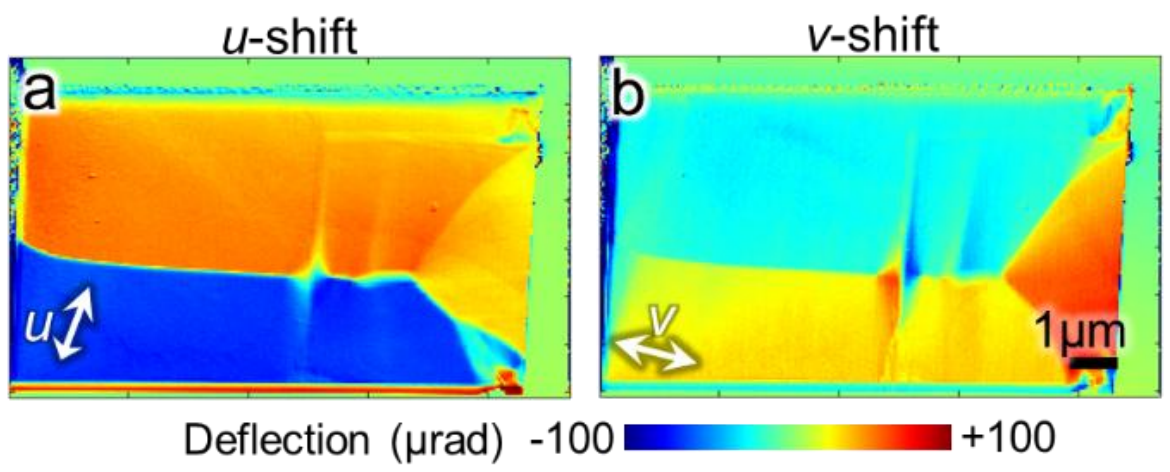

**Fig. 7** Edge detection DPC results. (a) Shift in ***u*** direction, and (b) shift in ***v*** direction.

## E. Quantitative Comparison

To more quantitatively compare the various proposed methods above rather than visual impression, we here evaluate the standard deviation of the DPC signal at a given area. As shown in Fig. 8a as a boxed area, we chose the evaluation region, where the diffraction contrast is severe while the magnetic contrast is relatively homogeneous (see magnetization maps in Fig. 6). This area corresponds to the position discussed in the previous section, which is indicated by the arrow in Fig. 6a, c. The evaluated mean beam shift magnitudes and angles within the selected area are shown in Fig. 8b-e for the CoM and CC methods as well as the edge detection. We include the edge detection results in CC in panels d and e as they are similar in principle. The standard deviations are indicated by error bars. Clearly, the CC approaches have less fluctuations in the mean values as well as in the standard deviations (errors), compared to the CoM methods. The edge detection approach further improves the signal. The deviation in the angle is especially significant for the single disk CoM signals. These fluctuations can be cancelled out by averaging multiple disk signals, ensuring that such a multiple angle illumination is effective. We also note that the CoM shift values are systematically smaller than CC values, which can be attributed to the finite background signals in the diffraction pattern. Overall it is clear that the "Entire CC" method (or the edge-detection method) gives the best signal.

The signal enhancement in the CC approach with multiple probes can also be quantitatively considered: By distributing the intensity into $n$ disks, the edge shift signal is enhanced by $\sqrt{n}$, given that the total area (beam current) is kept the same. Increasing the number of holes with random orientations also suppresses the directionality of the tilt patterns (see Figure S5 in the Supplementary Material), making the method more robust to the systematic diffractions in various directions.

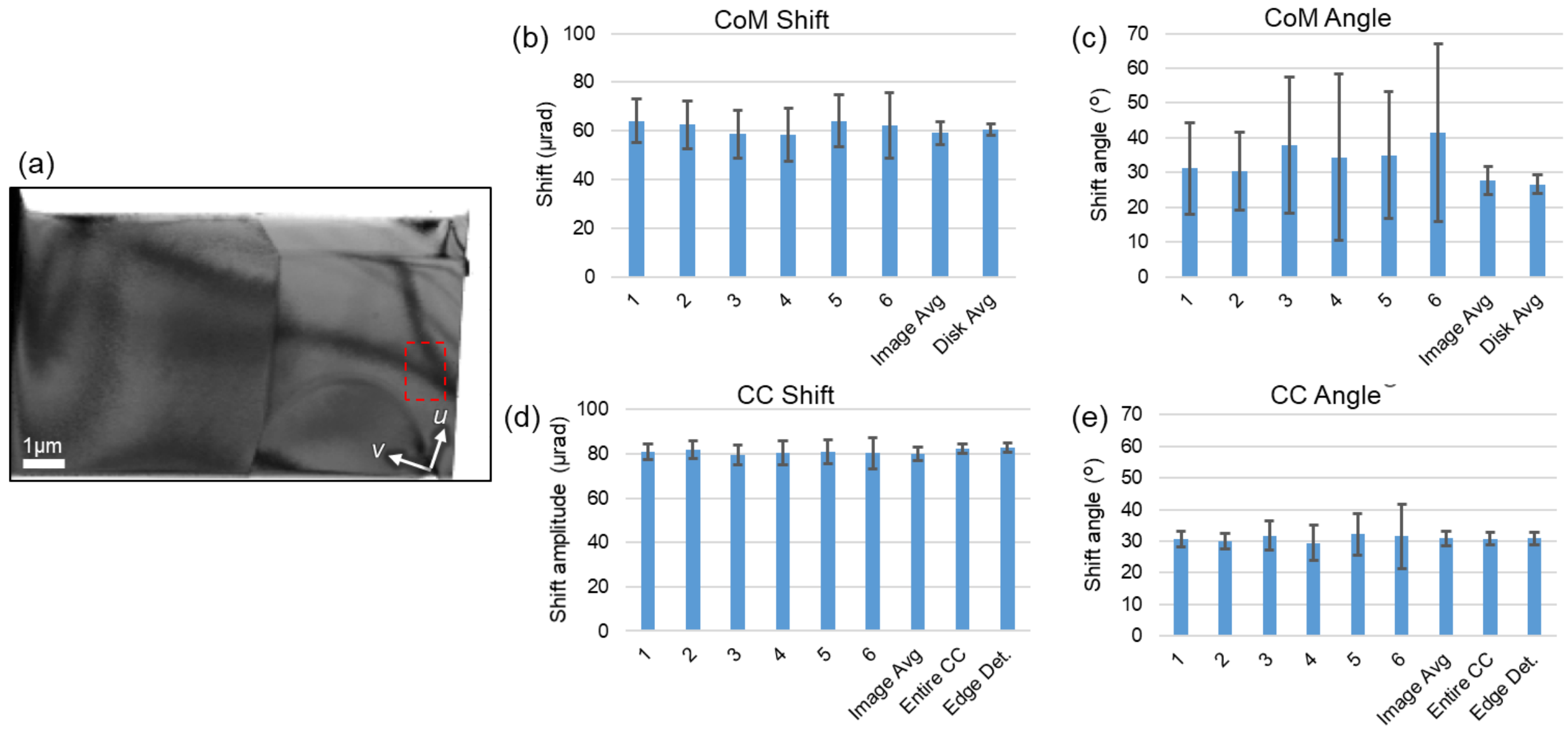

**Fig. 8** Comparison of the proposed methods. (a) Evaluation area indicated by a red dashed box (840 pixels) in the "all-disk" BF image, where the diffraction contrast is significant. (b, c) CoM evaluation of the boxed region for (b) mean beam shift magnitudes and (c) mean angles with respect to the ***v*** axis direction, with the standard deviation indicated by the error bars. The numbers indicated along the horizontal axis correspond to the single disk evaluation for the disks $i$ = 1-6. "Image Avg" and "Disk Avg" correspond to the CoM map image averaging (Fig. 4a, b) and disk averaging (Fig. 4c, d), respectively. (d, e) CC evaluation for (d) the shift magnitude and (e) its angle for each method. "Image Avg" and "Entire CC" correspond to the CC map image averaging (Fig. 6a, b) and CC map of the entire pattern (Fig. 6c, d), respectively. The edge detection results are also shown in the panels d and e together with the CC results.

## F. Magnetization mapping from entire-pattern CC

The entire-pattern CC images of Fig. 6c and d suppress the diffraction contrast almost completely and provide pure magnetic information while ensuring the signal to noise ratio. The magnetization vector maps, which are shown in Fig. 8, were reconstructed by combining the horizontal (***u***) and vertical (***v***) beam-shift components obtained from the entire-pattern analysis (Fig 6c, d). Colors and white vectors represent the direction and amplitude of the magnetic flux, with a color wheel provided in Fig. 9b. A closure-type magnetic domain structure with clockwise rotation is clearly visualized in Fig. 9a, which reflects the tendency of soft magnets to minimize magneto static energy by not exposing the magnetic poles. In conjunction with this, due to the exchange interaction and material anisotropy, magnetic domains with straight domain walls are formed so that the magnetic flux is closed. Fig. 9b presents an enlarged view of the boxed region in Fig. 9a, where the magnetization rotates around the 90° domain walls, labeled as B-B' and C-C', while the in-plane vector length becomes smaller at the 180° domain walls labeled as A-A'. They are consistent with Nèel-type and Bloch-type domain walls, respectively.

It is interesting that the horizontal 180° domain wall, which extends from the left until the violet domain, is partially broken around the center of Fig. 9a or at the left of Fig. 9b. And at this position, vertical 180° domain wall is partially formed. By comparing with the BF image of Fig. 2a, this can be attributed to the grain boundary or some defect where the exchange interaction is possibly smaller. Also it is noticeable that 180° domain wall is thicker on the right side of Fig. 9b, which also seems corresponding to another grain boundary according to the BF image of Fig. 2a. We here emphasize that such detailed magnetization analysis is possible because the diffraction contrasts are effectively removed by multi-angle beam and CC approach.

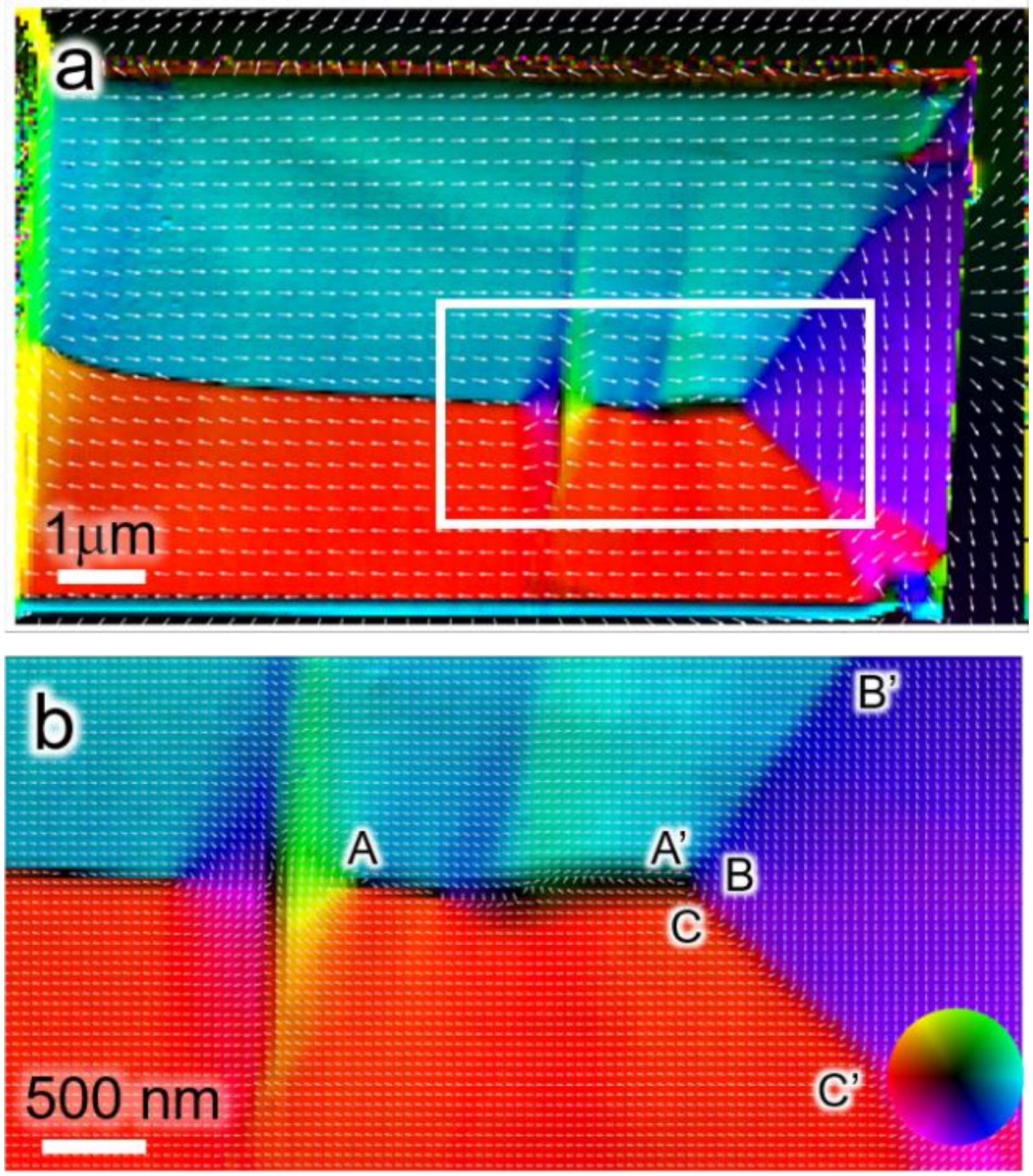

**Fig. 9.** Magnetization vector maps reproduced from the CC of the entire angular patterns (Fig. 6c, d). (a) The magnetization distribution of the entire field of view. (b) Enlarged view of the region marked by the white rectangle box in (a). The color map, with its color wheel shown at the bottom right of panel b, represents the orientation and the magnitude of the magnetization vectors. The white arrows indicate the direction of the magnetization.

## G. Extended information extraction from 4D-STEM datasets

To further analyze the sample, we here extract more information from the 4D dataset rather than simple DPC images. In the CoM analysis of Fig. 3 and 4, we discussed that the CoM shift is different for each disk when the diffraction contrast overlaps. Inversely, the deviations of the CoM signal for each disk should give more dominant diffraction contrast components. In Fig. 10a we plotted the standard deviation of each pixel from the six COM shift maps of Fig. 3 ( $\sigma_{\mathrm{CoM}}$). The bend contours of the specimen emerge clearly, especially where strong diffraction contrasts appear in the maps of Fig. 3. This way, one can confirm that the contrast is of diffraction effect, rather than the magnetic one. Similar standard deviation analysis can also be performed for the intensity of each disk, which correspond to the BF image series of Fig. S2 in the supplementary material. The intensity standard deviation map of Fig. 10b ($\sigma_{\mathrm{Intensity}}$) shows similar contours to the panel a. Comparing these two patterns, the narrow contour, corresponding to low index diffractions, are more enhanced in the $\sigma_{\mathrm{CoM}}$ map.

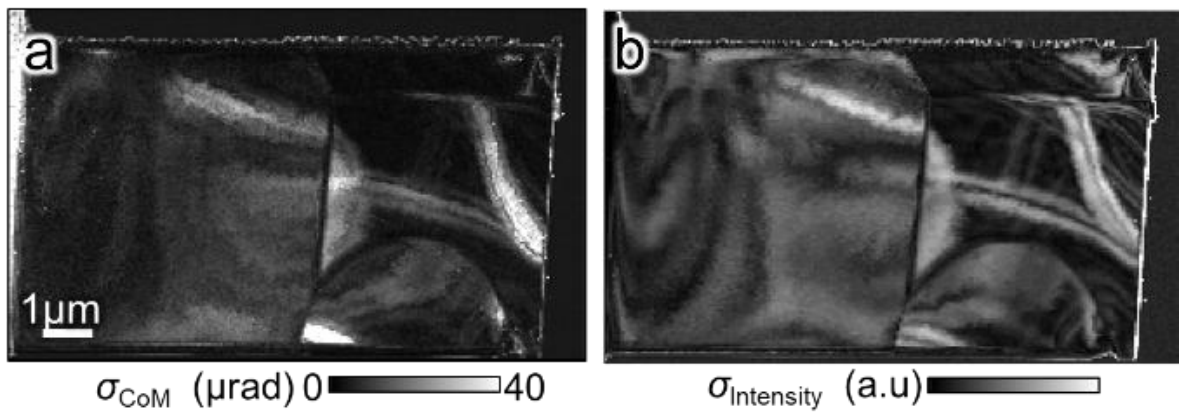

**Fig. 10.** Diffraction contrast analysis. (a) Standard deviation map of the CoM shift of six disks at each beam position, corresponding to the deviations of CoM maps of Fig. 3. (b) Standard deviation map of the intensity of six disks.

For the CC analysis, one can also extract more information than the CC peak shift. Here we utilize the aspect ratio of the CC peak of each pixel, as shown in Fig. 11. The aspect ratio amplitude map of Fig. 11a shows strongly enhanced signals at the 180° domain walls. This high signal of the CC peak aspect ratio corresponds to the deformation of the disk in the angular pattern, as shown in Fig. 11b and c for the 180° domain wall positions indicated in panel a. It is noted that the elongation direction of the CC spot, which is shown in the inset, is opposite to that of the disk. The deformation of the disk indicates that the electron beam has experienced not only simple lateral deflection (homogeneous phase inclination) but also more complicated phase changes within the beam diameter, which can be estimated to be ~ 5 nm from the angular size of each aperture disk. At the 180° domain wall positions, the magnetization varies abruptly in space, which results in the deformation of the disk and the elongation of the corresponding CC pattern. To map this CC peak features, the CC peak elongation direction is plotted in a color code in Fig. 11d. The amplitude is coded as the brightness of the color. In this representation (Fig. 11d), the deformation direction depends on the domain wall directions. The 90° domain wall with a green color at the bottom right of the image becomes also discernible because of the different disk elongation orientation from the surroundings. Since this disk deformation corresponds to the sub-beam size variation of the magnetic structure, the domain wall within the beam size could possibly be investigated. These analysis approaches highlight the potential of MaB-STEM combined with 4D datasets for advanced magnetic characterization and structural correlation.

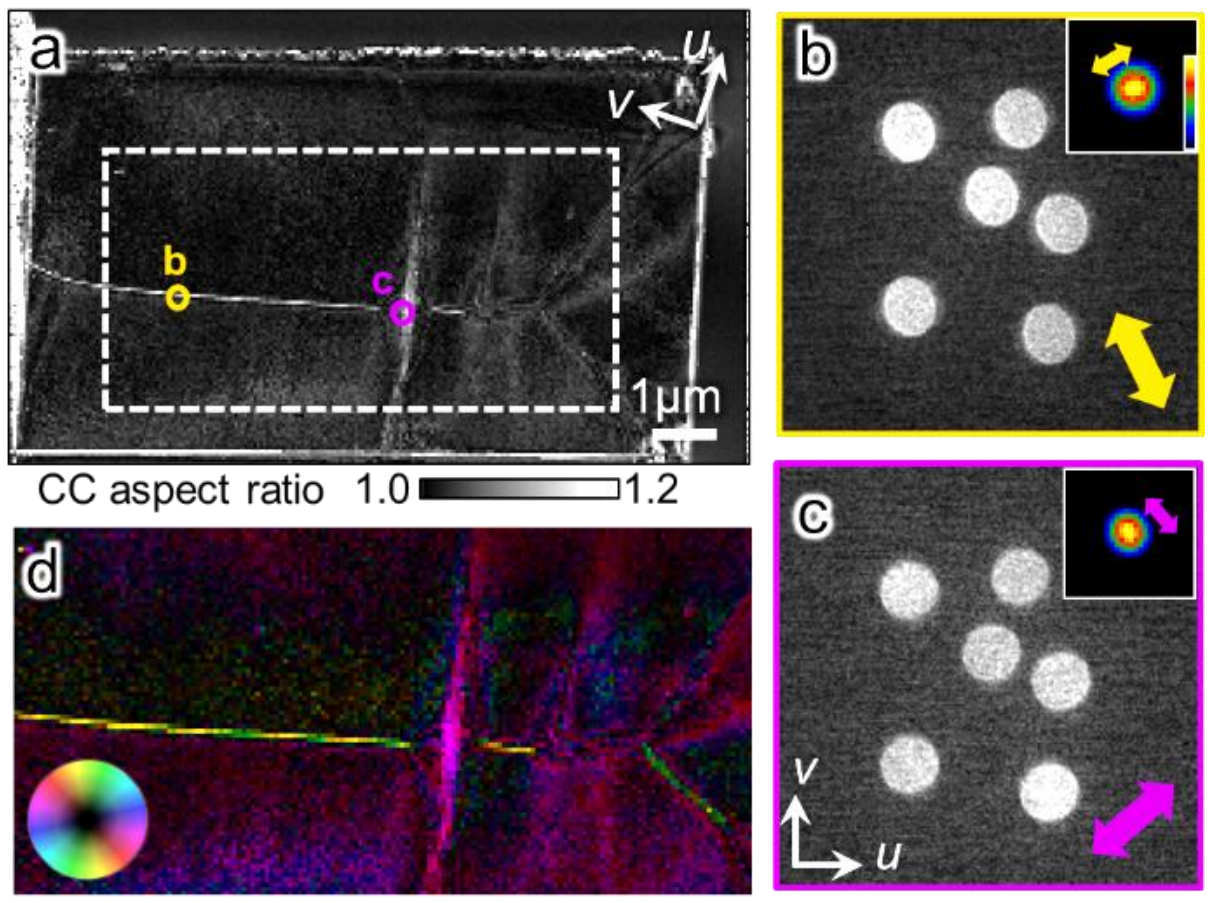

**Fig. 11.** Analysis of the angular disk deformation using CC peaks. (a) Aspect ratio map of the CC peak. (b, c) Representative angular patterns extracted from the positions indicated in panel a, where the deformation of the disk is significant. The inset shows the CC peak. The elongation direction of the disk patterns and CC peaks are shown by arrows. (d) Magnified aspect ratio map of the CC peak of the indicated area in panel a by dashed rectangle. The long axis direction and aspect ratio of the CC peak are displayed by colors and brightness, as shown in the color wheel.

## IV. CONCLUSION

We have demonstrated multi-angle beam scanning transmission electron microscopy (MaB-STEM) to visualize magnetization distribution, which significantly suppresses diffraction contrasts. This method leverages multi-beam illumination with different angles generated by a randomly distributed multi-hole aperture combined with differential phase contrast (DPC) imaging within the 4D-STEM framework. The diffraction contrasts originating from crystalline samples in DPC images can be reduced by image-processing multiple illumination disks in the angular pattern through center-of-mass (CoM) or cross-correlation (CC) analysis. Especially, the CC method can reproduce magnetization mappings almost completely free from diffraction contrasts. Furthermore, 4D data analysis allows evaluating the influence of the diffraction contrasts, as well as abrupt magnetization changes even within the size of the beam spot (spatial resolution) in the real space, as demonstrated by aspect ratio plots of the CC peaks. Such 4D-STEM approaches with optimized aperture shapes possess significant potentials as next-generation STEM methods, possibly beyond the demonstrated magnetic imaging. Its ability to extract extended information from 4D-STEM datasets could be applied e.g. to advanced crystalline specimen characterization, strain mapping, defect analysis etc.

**Supplementary material.**

See the supplementary material for additional figures, experimental conditions, and analysis. It includes disk-resolved BF images, schematic explanations of the CoM and CC methods, comparison of different averaging approaches, and discussions on the influence of aperture geometry on diffraction-contrast suppression.

## ACKNOWLEDGEMENTS

This work is supported by JSPS Kakenhi (JP24H00400), JST CREST (JPMJCR25I3), and JST FOREST (JPMJFR213J).

## AUTHOR DECLARATIONS

### Conflict of Interest

The authors have no conflicts to disclose.

## Author Contributions

**Akira Yasuhara** : Data curation (equal); Resources (lead); Project administration (equal); Visualization (supporting); Writing – original draft (equal); Writing – review & editing (equal). **Zentaro Akase** : Resources (equal); Validation (supporting); Writing – review & editing (supporting). **Takumi Sannomiya :** Conceptualization (lead); Data curation (lead); Formal analysis(lead); Funding acquisition (lead); Investigation (lead); Methodology (lead); Project administration (lead); Validation (lead); Visualization (lead); Writing – original draft (equal); Writing – review & editing(equal).

## DATA AVAILABILTY

Data will be made available on request.

# Supplementary Material

# Multi-Angle Beam Scanning Transmission Electron Microscopy for Diffraction-Contrast-Free Magnetic Imaging

*Akira Yasuhara*[1], Zentaro Akase[2] and Takumi Sannomiya*[3]*

[1] JEOL Ltd. 3-1-2 Musashino, Akishima, Tokyo, 196-8558, Japan.

[2] Division of Materials Science, Nara Institute of Science and Technology, 8916-5 Takayama, Ikoma, Nara 630-0192, Japan

[3] Department of Materials Science and Engineering, School of Materials and Chemical Technologies, Institute of Science Tokyo, 4259 Nagatsuta, Midoriku, Yokohama, 226-8501, Japan.

**Corresponding Authors**

* Akira Yasuhara (Email: ayasuhar@jeol.co.jp)

* Takumi Sannomiya (Email: sannomiya@mct.isct.ac.jp)

**Table S1.** Additional experimental conditions

| Microscope experimental condition | |
|---|---|
| accelaration voltage | 200 kV |
| convergence semi-angle of each probe | 250 μrad |
| ronchigram flat region (radius) | 3 mrad |
| maximu beam tilt angle | ±2.5 mrad |
| probe current of each electorn beam | 15 pA |
| camera length | 1500 mm |
| **4D-STEM condition** | |
| detector | Oneview CMOS camera |
| dwell time | 50 msec/pixel |
| image pixel size | 50 nm/pixel |
| detector sampling | 256×256 pixels (4× binning) |
| total acquisiton time | less than 3 minites |

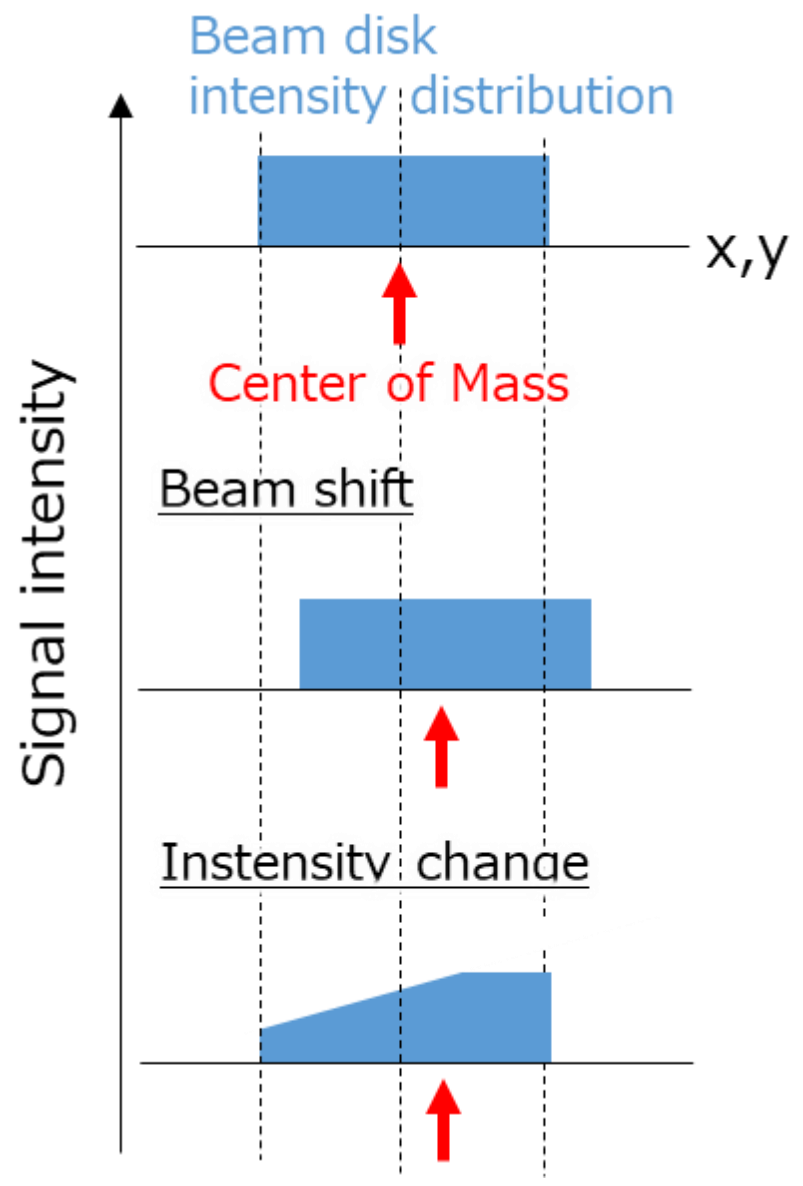

**Figure S1.** Schematic illustration of center of mass (CoM) evaluation and the influence of the intensity distribution change by diffraction contrast.

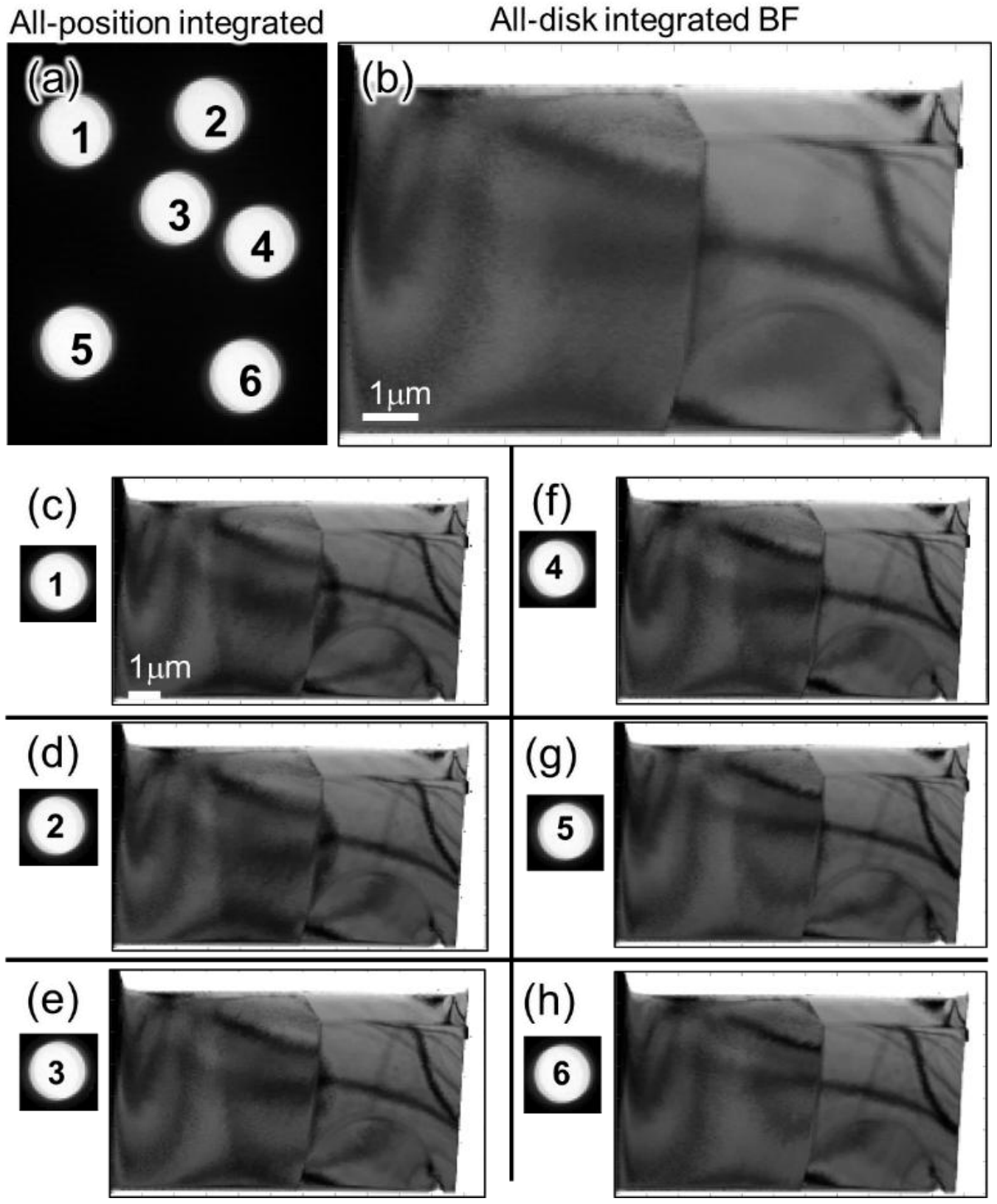

**Figure S2.** (a) Position-integrated angular pattern from the 4D-STEM dataset. (b) STEM bright-field (BF) image reconstructed by integrating the intensity of the entire angular pattern. (c-h) BF-STEM images using individual disks #1–6, as indicated in panel a. The disk patterns are extracted as circular regions with a diameter of 36 pixels. Each BF image was reconstructed by integrating the total intensity from the corresponding disks.

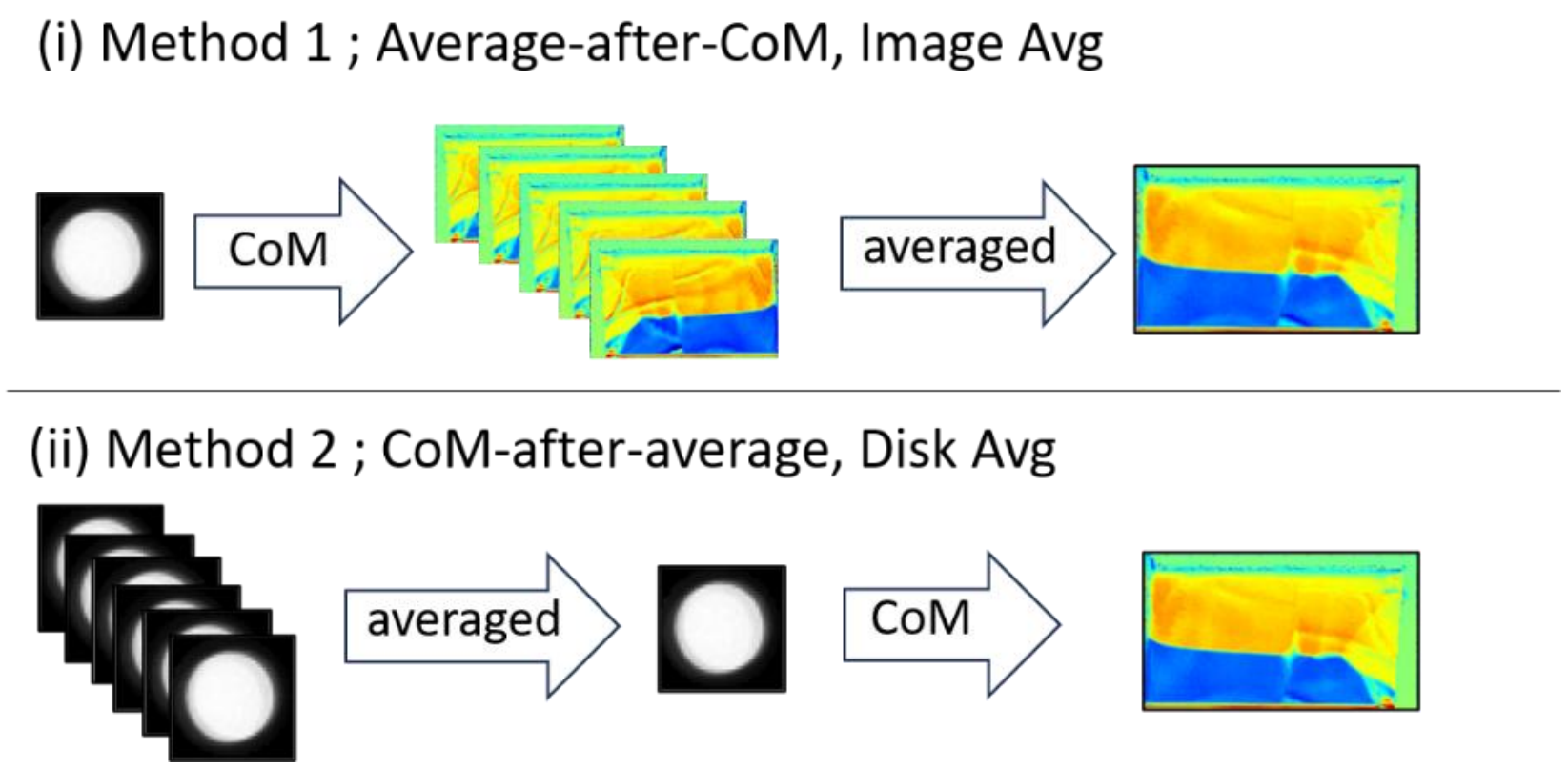

**Figure S3.** Schematic comparison of two CoM-based averaging approaches used for visualization of magnetic field. (i) Averaged-after-CoM approach. Individual CoM image are first calculated from each discs, and the resulting images are subsequently averaged. (ii) CoM-after-average approach. The six aperture disks are first superposed and averaged, and the CoM image is then calculated from the averaged diffraction pattern.

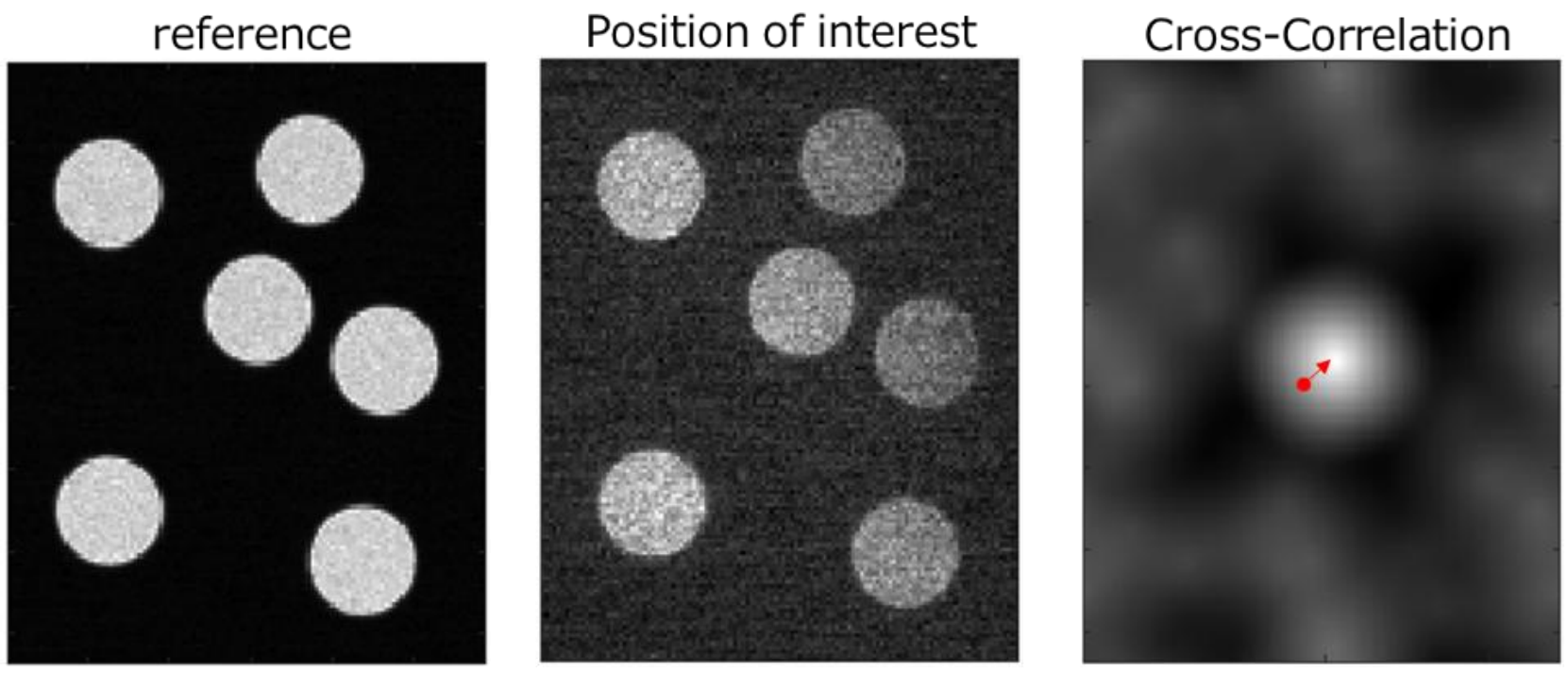

**Figure S4.** Cross correlation process. The cross-correlation (CC) pattern (right most) is generated by the diffraction pattern at each beam position (middle) with the reference (left). The CC peak is fitted by a paraboloid function to determine the peak shift vector, which is schematically indicated by the red arrow.

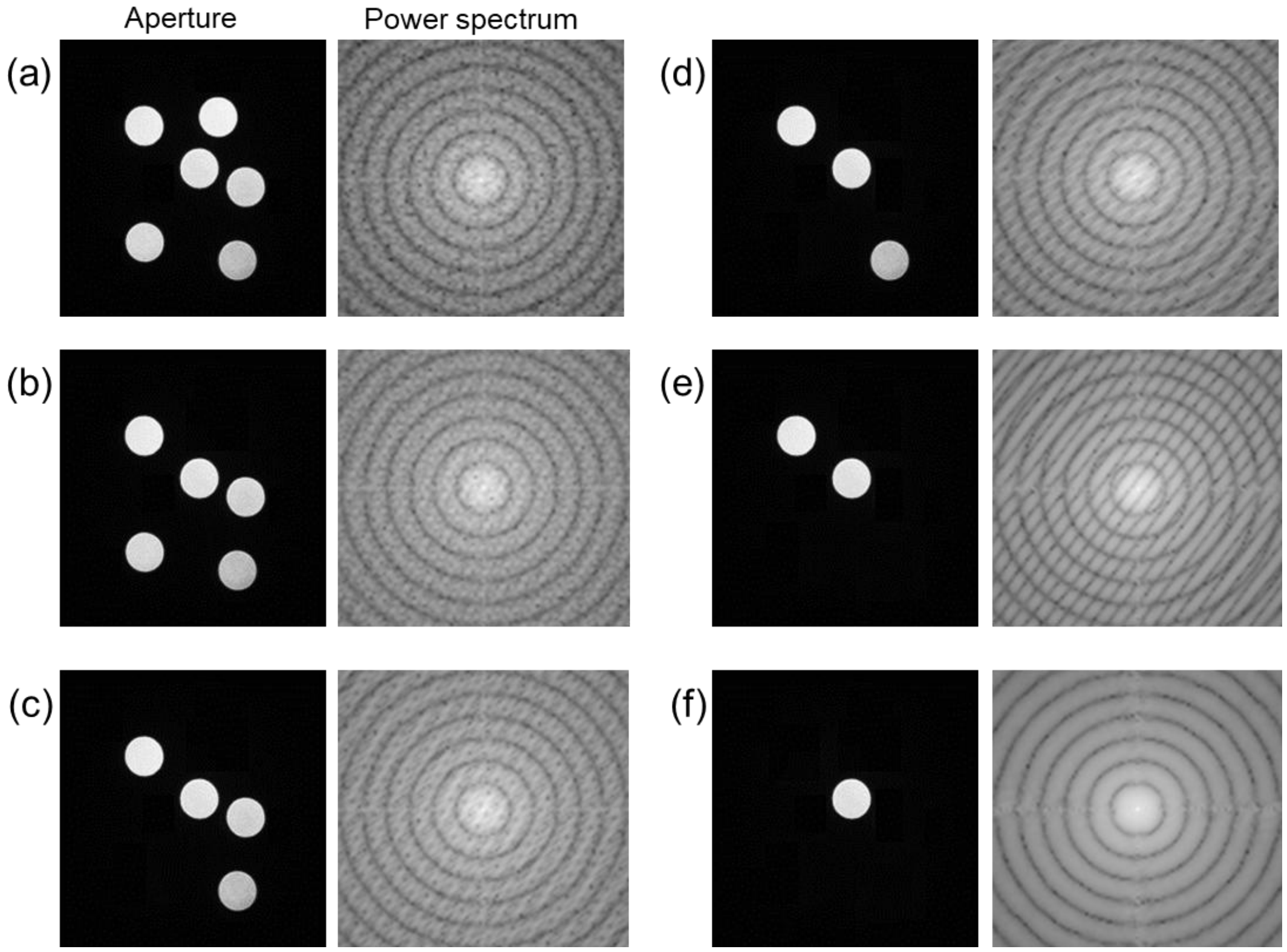

**Figure S5.** Power spectra (corresponding to the beam shape) of aperture patterns with different number of holes. With too few holes (e.g. panels c-e), one can see direction interference patterns indicating the directionality, which can be sensitive to specific diffraction directions. With six randomly oriented holes (panel a), such directionality is suppressed.